\documentclass[11pt]{article}

\usepackage{epsfig}
\usepackage{amssymb}
\usepackage{rotating}

\begin{document}         

%
%
\newcommand{\EPnum}     {CERN-EP/2001-082}
\newcommand{\PRnum}     {OPAL Paper PR-349}
\newcommand{\TNnum}     {OPAL Technical Note TN-xxx}
\newcommand{\Date}      {November 16th, 2001}
\newcommand{\Author}    {C.~Y.~Chang, M.~Gruw\'e and G.~A.~Snow}
\newcommand{\MailAddr}  {magali.gruwe@cern.ch}
\newcommand{\EdBoard}   {P.~Bock, D.~Horvath, \\
                         A.~Okpara and E.~Torrence}
\newcommand{\DraftVer}  {Version 1.0}
\newcommand{\DraftDate} {\today}
\newcommand{\TimeLimit} {Friday, November 9th, 16:00}


\newcommand{\ellp}      {\ell^+}
\newcommand{\ellm}      {\ell^-}
\newcommand{\Zboson}    {{\mathrm Z}^{0}}
\newcommand{\Wpm}       {{\mathrm W}^{\pm}}

\newcommand{\spar}      {\tilde{\mathrm s}}

\newcommand{\gluino}    {\tilde{\mathrm g}}

\newcommand{\sfer}      {\tilde{\mathrm f}}
\newcommand{\seff}      {\tilde{\mathrm f}}

\newcommand{\squark}    {\tilde{q}}
\newcommand{\sq}        {\tilde{\mathrm q}}
\newcommand{\stopm}     {\tilde{\mathrm{t}}_{1}}
\newcommand{\stopn}     {\tilde{\mathrm{t}}}
\newcommand{\stops}     {\tilde{\mathrm{t}}_{2}}
\newcommand{\stopbar}   {\bar{\tilde{\mathrm{t}}}_{1}}
\newcommand{\stopx}     {\tilde{\mathrm{t}}}
\newcommand{\stopl}     {\tilde{\mathrm{t}}_{\mathrm L}}
\newcommand{\stopr}     {\tilde{\mathrm{t}}_{\mathrm R}}

\newcommand{\slepton}   {\tilde{\ell}}
\newcommand{\sele}      {\tilde{\mathrm e}}
\newcommand{\sell}      {\tilde{\ell}}
\newcommand{\smu}       {\tilde{\mu}}
\newcommand{\stau}      {\tilde{\tau}}
\newcommand{\sneutrino} {\tilde{\nu}}
\newcommand{\snu}       {\tilde{\nu}}

\newcommand{\chp}       {\tilde{\chi}^+_1}
\newcommand{\chip}      {\tilde{\chi}^+_1}
\newcommand{\chim}      {\tilde{\chi}^-_1}
\newcommand{\chpm}      {\tilde{\chi}^\pm_1}
\newcommand{\chipm}     {\tilde{\chi}^\pm_1}
\newcommand{\nt}        {\tilde{\chi}^0}
\newcommand{\chin}      {\tilde{\chi }^{0}_{1}}
\newcommand{\neutralino}{\tilde{\chi }^{0}_{1}}
\newcommand{\neutrala}  {\tilde{\chi }^{0}_{2}}
\newcommand{\neutralb}  {\tilde{\chi }^{0}_{3}}
\newcommand{\neutralc}  {\tilde{\chi }^{0}_{4}}
\newcommand{\bino}      {\tilde{\mathrm B}^{0}}
\newcommand{\wino}      {\tilde{\mathrm W}^{0}}
\newcommand{\higginoa}  {\tilde{\rm H_{1}}^{0}}
\newcommand{\higginob}  {\tilde{\mathrm H_{1}}^{0}}
\newcommand{\chargino}  {\tilde{\chi }^{\pm}_{1}}
\newcommand{\charginop} {\tilde{\chi }^{+}_{1}}

\newcommand{\mdbHpp}    {\mathrm{H}^{++}}
\newcommand{\mdbHmm}    {\mathrm{H}^{--}}
\newcommand{\mdbHpmpm}  {\mathrm{H}^{\pm\pm}}


\newcommand{\el}        {\mbox{${\mathrm e}^-$}}
\newcommand{\lept}      {\mbox{$\ell^-$}}
\newcommand{\nul}       {\mbox{$\nu_\ell$}}
\newcommand{\nubar}     {\mbox{$\overline{\nu}_\ell$}}
\newcommand{\w}         {\mbox{W$^\pm$}}
\newcommand{\Wstar}     {W$^{(*)}$}

\newcommand{\selem}     {\mbox{$\tilde{\mathrm e}^-$}}
\newcommand{\smum}      {\mbox{$\tilde\mu^-$}}
\newcommand{\staum}     {\mbox{$\tilde\tau^-$}}
\newcommand{\slept}     {\mbox{$\tilde{\ell}^\pm$}}
\newcommand{\sleptm}    {\mbox{$\tilde{\ell}^-$}}
\newcommand{\ch}        {\mbox{$\tilde{\chi}^\pm_1$}}
\newcommand{\chm}       {\mbox{$\tilde{\chi}^-_1$}}
\newcommand{\chmp}      {\mbox{$\tilde{\chi}^\pm_1$}}
\newcommand{\chz}       {\mbox{$\tilde{\chi}^0_1$}}

\newcommand{\Hl}        {\mbox{$\mathrm{L}^\pm$}}
\newcommand{\Hm}        {\mbox{$\mathrm{L}^-$}}
\newcommand{\Hnu}       {\mbox{$\nu_{\mathrm{L}}$}}

\newcommand{\dbHpp}     {\mbox{$\mathrm{H}^{++}$}}
\newcommand{\dbHmm}     {\mbox{$\mathrm{H}^{--}$}}
\newcommand{\dbHpmpm}   {\mbox{$\mathrm{H}^{\pm\pm}$}}
\newcommand{\dbHLpp}    {\mbox{$\mathrm{H}^{++}_{\mathrm{L}}$}}
\newcommand{\dbHLmm}    {\mbox{$\mathrm{H}^{--}_{\mathrm{L}}$}}
\newcommand{\dbHLpmpm}  {\mbox{$\mathrm{H}^{\pm\pm}_{\mathrm{L}}$}}
\newcommand{\dbHRpp}    {\mbox{$\mathrm{H}^{++}_{\mathrm{R}}$}}
\newcommand{\dbHRmm}    {\mbox{$\mathrm{H}^{--}_{\mathrm{R}}$}}
\newcommand{\dbHRpmpm}  {\mbox{$\mathrm{H}^{\pm\pm}_{\mathrm{R}}$}}
\newcommand{\MZ}        {M_{\mathrm Z}}
\newcommand{\Mt}        {M_{\mathrm t}}
\newcommand{\mscalar}   {m_{0}}
\newcommand{\Mgaugino}  {M_{1/2}}
\newcommand{\rs}        {\sqrt{s}}
\newcommand{\MGUT}      {M_{\mathrm {GUT}}}
\newcommand{\mixang}    {\theta _{\mathrm {mix}}}
\newcommand{\mstop}     {m_{\stopm}}
\newcommand{\msell}     {m_{\sell}}
\newcommand{\mchi}      {m_{\neutralino}}

\newcommand{\Rparity}   {$R$-parity}
\newcommand{\Rp}        {$R_{p}$}
\newcommand{\mch}       {\mbox{$m_{\tilde{\chi}^\pm_1}$}}
\newcommand{\mslept}    {\mbox{$m_{\tilde{\ell}}$}}
\newcommand{\mstau}     {\mbox{$m_{\staum}$}}
\newcommand{\msmu}      {\mbox{$m_{\smum}$}}
\newcommand{\msele}     {\mbox{$m_{\selem}$}}
\newcommand{\mchz}      {\mbox{$m_{\tilde{\chi}^0_1}$}}
\newcommand{\dm}        {\mbox{$\Delta m$}}
\newcommand{\dmch}      {\mbox{$\Delta m_{\ch-\chz}$}}
\newcommand{\dmslept}   {\mbox{$\Delta m_{\slept-\chz}$}}
\newcommand{\dmhl}      {\mbox{$\Delta m_{\Hl-\Hnu}$}}

\newcommand{\ee}        {{\mathrm e}^+ {\mathrm e}^-}
\newcommand{\qq}        {{\mathrm q}\bar{\mathrm q}}
\newcommand{\nunu}      {\nu \bar{\nu}}
\newcommand{\mumu}      {\mu^+ \mu^-}
\newcommand{\tautau}    {\tau^+ \tau^-}
\newcommand{\ellell}    {\ell^+ \ell^-}
\newcommand{\nulqq}     {\nu \ell {\mathrm q} \bar{\mathrm q}'}
\newcommand{\ff}        {{\mathrm f} \bar{\mathrm f}}
\newcommand{\WW}        {{\mathrm W}^+{\mathrm W}^-}
\newcommand{\allqq}     {\sum_{q \neq t} q \bar{q}}
\newcommand{\sleppair}  {\sell^+ \sell^-}

\newcommand{\pp}        {p \bar{p}}
\newcommand{\epair}     {\mbox{${\mathrm e}^+{\mathrm e}^-$}}
\newcommand{\mupair}    {\mbox{$\mu^+\mu^-$}}
\newcommand{\taupair}   {\mbox{$\tau^+\tau^-$}}
\newcommand{\qpair}     {\mbox{${\mathrm q}\overline{\mathrm q}$}}
\newcommand{\eeee}      {\mbox{\epair\epair}}
\newcommand{\eemumu}    {\mbox{\epair\mupair}}
\newcommand{\eetautau}  {\mbox{\epair\taupair}}
\newcommand{\eeqq}      {\mbox{\epair\qpair}}
\newcommand{\fs}        { final states}
\newcommand{\epairf}    {\mbox{\epair\fs}}
\newcommand{\mupairf}   {\mbox{\mupair\fs}}
\newcommand{\taupairf}  {\mbox{\taupair\fs}}
\newcommand{\qpairf}    {\mbox{\qpair\fs}}
\newcommand{\eeeef}     {\mbox{\eeee\fs}}
\newcommand{\eemumuf}   {\mbox{\eemumu\fs}}
\newcommand{\eetautauf} {\mbox{\eetautau\fs}}
\newcommand{\eeqqf}     {\mbox{\eeqq\fs}}
\newcommand{\ffff}      {four fermion final states}
\newcommand{\llnunu}    {\mbox{\lpair\nul\nubar}}
\newcommand{\lnuqq}     {\mbox{\lept\nubar\qpair}}
\newcommand{\eeeell}    {\mbox{\epair$\rightarrow$\epair\lpair}}
\newcommand{\eell}      {\mbox{\epair\lpair}}
\newcommand{\llgam}     {\mbox{$\ell\ell(\gamma)$}}
\newcommand{\nunugam}   {\mbox{$\nu\bar{\nu}\gamma\gamma$}}
\newcommand{\zee}       {\mbox{Zee}}
\newcommand{\zzg}       {\mbox{ZZ/Z$\gamma$}}
\newcommand{\wenu}      {\mbox{We$\nu$}}
\newcommand{\lpair}     {\mbox{$\ell^+\ell^-$}}
\newcommand{\wpair}     {\mbox{$W^+W^-$}}
\newcommand{\spair}     {\mbox{$\tilde{\ell}^+\tilde{\ell}^-$}}
\newcommand{\staupair}  {\mbox{$\tilde{\tau}^+\tilde{\tau}^-$}}
\newcommand{\smupair}   {\mbox{$\tilde{\mu}^+\tilde{\mu}^-$}}
\newcommand{\selepair}  {\mbox{$\tilde{\mathrm e}^+\tilde{\mathrm e}^-$}}
\newcommand{\chpair}    {\mbox{$\tilde{\chi}^+_1\tilde{\chi}^-_1$}}
\newcommand{\dch}       {\mbox{\chm$\rightarrow$\chz\lept\nubar}}
\newcommand{\dslept}    {\mbox{\sleptm$\rightarrow$\chz\lept}}
\newcommand{\dH}        {\mbox{\Hm$\rightarrow$\lept\nubar\Hnu}}
\newcommand{\dW}        {\mbox{W$^-\rightarrow$\lept\nubar}}
\newcommand{\dsele}     {\mbox{\selem$\rightarrow$\chz e$^-$}}

\newcommand{\thrust}    {T}
\newcommand{\nthrust}   {\hat{n}_{\mathrm{thrust}}}
\newcommand{\thethr}    {\theta_{\,\mathrm{thrust}}}
\newcommand{\phithr}    {\phi_{\mathrm{thrust}}}
\newcommand{\acosthr}   {|\cos\thethr|}
\newcommand{\thejet}    {\theta_{\,\mathrm{jet}}}
\newcommand{\acosjet}   {|\cos\thejet|}
\newcommand{\thmiss}    { \theta_{\mathrm{miss}} }
\newcommand{\cosmiss}   {| \cos \thmiss |}
\newcommand{\phiacop}   {\phi_{\mathrm{acop}}}
\newcommand{\thacop}    {\theta _{\mathrm {Acop}}}
\newcommand{\cosjet}    {\cos\thejet}
\newcommand{\costhr}    {\cos\thethr}
\newcommand{\djoin}     {d_{\mathrm{join}}}
\newcommand{\plept}     {p_{\mathrm{lept}}}
\newcommand{\nlept}     {N_{\mathrm{lept}}}
\newcommand{\ntracks}   {N_{\mathrm{tracks}}}

\newcommand{\roots}     {\sqrt{s}}
\newcommand{\Evis}      {E_{\mathrm{vis}}}
\newcommand{\Rvis}      {E_{\mathrm{vis}}\,/\roots}
\newcommand{\Mvis}      {m_{\mathrm{vis}}}
\newcommand{\Rbal}      {R_{\mathrm{bal}}}

\newcommand{\acopc}     {\mbox{$\phi^{\mathrm{acop}}$}}
\newcommand{\acolc}     {\mbox{$\theta^{\mathrm{acol}}$}}
\newcommand{\acop}      {\mbox{$\phi_{\mathrm{acop}}$}}
\newcommand{\acol}      {\mbox{$\theta_{\mathrm{acol}}$}}
\newcommand{\pt}        {\mbox{$p_{\mathrm{t}}$}}
\newcommand{\pz}        {\mbox{$p_{\mathrm{z}}^{\mathrm{miss}}$}}
\newcommand{\ptevt}     {\mbox{$p_{t}^{\mathrm{miss}}$}}
\newcommand{\ptaxic}    {\mbox{$a_{t}^{\mathrm{miss}}$}}
\newcommand{\stevt}     {\mbox{$p_{t}^{\mathrm{miss}}$/\Ebeam}}
\newcommand{\staxic}    {\mbox{$a_{t}^{\mathrm{miss}}$/\Ebeam}}
\newcommand{\dptaxic}   {\mbox{missing $p_{t}$ wrt. event axis \ptaxic}}
\newcommand{\cosevt}    {\mbox{$\mid\cos\theta_{\mathrm{p}}^{\mathrm{miss}}\mid$}}
\newcommand{\axicos}    {\mbox{$\mid\cos\theta_{\mathrm{a}}^{\mathrm{miss}}\mid$}}
\newcommand{\pthet}     {\mbox{$\theta_{\mathrm{p}}^{\mathrm{miss}}$}}
\newcommand{\athet}     {\mbox{$\theta_{\mathrm{a}}^{\mathrm{miss}}$}}
\newcommand{\dcosevt}   {\mbox{$\mid\cos\theta\mid$ of missing p$_{t}$}}
\newcommand{\daxicos}   {\mbox{$\mid\cos\theta\mid$ of missing p$_{t}$ wrt. event axis}}
\newcommand{\efdsw}     {\mbox{$x_{\mathrm{FDSW}}$}}
\newcommand{\acopf}     {\mbox{$\Delta\phi_{\mathrm{FDSW}}$}}
\newcommand{\acopm}     {\mbox{$\Delta\phi_{\mathrm{MUON}}$}}
\newcommand{\acopt}     {\mbox{$\Delta\phi_{\mathrm{trk}}$}}
\newcommand{\po}        {\mbox{$E_{\mathrm{isol}}^\gamma$}}
\newcommand{\qprod}     {\mbox{$q1$$*$$q2$}}
\newcommand{\lcode}     {lepton identification code}
\newcommand{\nctro}     {\mbox{$N_{\mathrm{trk}}^{\mathrm{out}}$}}
\newcommand{\necao}     {\mbox{$N_{\mathrm{ecal}}^{\mathrm{out}}$}}
\newcommand{\mout}      {\mbox{$m^{\mathrm{out}}$}}
\newcommand{\nctec}     {\mbox{\nctro$+$\necao}}
\newcommand{\gfract}    {\mbox{$F_{\mathrm{good}}$}}
\newcommand{\zz}        {\mbox{$|z_0|$}}
\newcommand{\dz}        {\mbox{$|d_0|$}}
\newcommand{\sint}      {\mbox{$\sin\theta$}}
\newcommand{\cost}      {\mbox{$\cos\theta$}}
\newcommand{\mcost}     {\mbox{$|\cos\theta|$}}
\newcommand{\dedx}      {\mbox{$dE/dx$}}
\newcommand{\wdedx}     {\mbox{$W_{dE/dx}$}}
\newcommand{\xe}        {\mbox{$x_E$}}
\newcommand{\ssix}      {\mbox{$\sqrt{s}$~=~161~GeV}}
\newcommand{\sthree}    {\mbox{$\sqrt{s}$~=~130--136~GeV}}
\newcommand{\mrecoil}   {\mbox{$m_{\mathrm{recoil}}$}}
\newcommand{\llmass}    {\mbox{$m_{ll}$}}
\newcommand{\acope}     {\mbox{$\Delta\phi_{\mathrm{EE}}$}}
\newcommand{\nee}       {\mbox{N$_{\mathrm{EE}}$}}
\newcommand{\eesum}     {\mbox{$\Sigma_{\mathrm{EE}}$}}
\newcommand{\at}        {\mbox{$a_{t}$}}
\newcommand{\spp}       {\mbox{$p$/\Ebeam}}
\newcommand{\acoph}     {\mbox{$\Delta\phi_{\mathrm{HCAL}}$}}

\newcommand{\Ecm}       {\mbox{$E_{\mathrm{cm}}$}}
\newcommand{\Ebeam}     {\mbox{$E_{\mathrm{beam}}$}}
\newcommand{\ipb}       {\mbox{pb$^{-1}$}}
\newcommand{\wrt}       {with respect to}
\newcommand{\sm}        {Standard Model}
\newcommand{\smb}       {Standard Model background}
\newcommand{\smp}       {Standard Model processes}
\newcommand{\smc}       {Standard Model Monte Carlo}
\newcommand{\mc}        {Monte Carlo}
\newcommand{\btb}       {back-to-back}
\newcommand{\tp}        {two-photon}
\newcommand{\tpb}       {two-photon background}
\newcommand{\tpp}       {two-photon processes}
\newcommand{\lp}        {lepton pairs}
\newcommand{\vto}       {\mbox{$\tau$ veto}}
\newcommand{\sml}       {\mbox{Standard Model \lpair$\nu\nu$ events}}
\newcommand{\sme}       {\mbox{Standard Model events}}
\newcommand{\sig}       {events containing a lepton pair plus missing transverse momentum}
\newcommand{\gsim}{\;\raisebox{-0.9ex}
           {$\textstyle\stackrel{\textstyle >}{\sim}$}\;}
\newcommand{\lsim}{\;\raisebox{-0.9ex}{$\textstyle\stackrel{\textstyle<}
           {\sim}$}\;}
\newcommand{\degree}    {^\circ}

%
%
\newcommand{\ZP}        {Z.~Phys.}
\newcommand{\PL}[3]     {Phys. Lett. {\bf B#1} (#2) #3.}
\newcommand{\etal}      {{\it et al}.,\,\ }
\newcommand{\PhysLett}  {Phys.~Lett.}
\newcommand{\PRL}       {Phys.~Rev.\ Lett.}
\newcommand{\PhysRep}   {Phys.~Rep.}
\newcommand{\PhysRev}   {Phys.~Rev.}
\newcommand{\NPhys}     {Nucl.~Phys.}
\newcommand{\NIM}       {Nucl.~Instr.\ Meth.}
\newcommand{\CPC}       {Comp.~Phys.\ Comm.}
\newcommand{\ZPhys}     {Z.~Phys.}
\newcommand{\IEEENS}    {IEEE Trans.\ Nucl.~Sci.}
\newcommand{\EuroPhys}  {Euro.~Phys. \ Jour.}
%
%
\newcommand{\OPALColl}  {OPAL Collab.}
\newcommand{\AColl}     {ALEPH Collab.}
\newcommand{\DColl}     {DELPHI Collab.}
\newcommand{\LColl}     {L3 Collab.}
\newcommand{\JADEColl}  {JADE Collab.}
%
\newcommand{\onecol}[2] {\multicolumn{1}{#1}{#2}}
\newcommand{\ra}        {\rightarrow}   


\begin{center}
{\Large EUROPEAN ORGANIZATION FOR NUCLEAR RESEARCH}
\end{center}

\begin{flushright}

\Large
    \EPnum\\
    \Date

\end{flushright}

\vspace{0.8cm}

\begin{center}

{    \huge \bf \boldmath

Search for Doubly Charged Higgs Bosons with the OPAL
detector at LEP }

\normalsize

\vspace{0.5cm}

\LARGE

The OPAL Collaboration

\end{center}

\vspace{1.0cm}

\begin{abstract}
A search for pair-produced doubly charged Higgs bosons 
has been performed 
using data samples corresponding to an integrated luminosity of
about 614~pb$^{-1}$ 
collected with the OPAL detector at LEP
at centre-of-mass energies between 189~GeV and 209~GeV. 
No evidence for a signal has been observed.
A mass limit of 98.5~GeV/$c^{2}$ at the 95\% confidence level has 
been set for 
the doubly charged Higgs particle in left-right symmetric models.
This is the first search for doubly charged Higgs bosons at centre-of-mass 
energies larger than 91~GeV.
\end{abstract}

\vspace{0.5cm}

\begin{center}

\vspace{2.5cm}

\Large \em To be submitted to Phys. Lett. B

\end{center}

\newpage


\begin{center}{\Large        The OPAL Collaboration
}\end{center}\bigskip
\begin{center}{
G.\thinspace Abbiendi$^{  2}$,
C.\thinspace Ainsley$^{  5}$,
P.F.\thinspace {\AA}kesson$^{  3}$,
G.\thinspace Alexander$^{ 22}$,
J.\thinspace Allison$^{ 16}$,
G.\thinspace Anagnostou$^{  1}$,
K.J.\thinspace Anderson$^{  9}$,
S.\thinspace Arcelli$^{ 17}$,
S.\thinspace Asai$^{ 23}$,
D.\thinspace Axen$^{ 27}$
% ...
}\end{center}\bigskip
\bigskip
$^{  1}$School of Physics and Astronomy, University of Birmingham,
Birmingham B15 2TT, UK
\newline
$^{  2}$Dipartimento di Fisica dell' Universit\`a di Bologna and INFN,
I-40126 Bologna, Italy
\newline
$^{  3}$Physikalisches Institut, Universit\"at Bonn,
D-53115 Bonn, Germany
\newline
% ...
$^{  5}$Cavendish Laboratory, Cambridge CB3 0HE, UK
\newline
% ...
$^{  9}$Enrico Fermi Institute and Department of Physics,
University of Chicago, Chicago IL 60637, USA
\newline
% ...
$^{ 16}$Department of Physics, Schuster Laboratory, The University,
Manchester M13 9PL, UK
\newline
$^{ 17}$Department of Physics, University of Maryland,
College Park, MD 20742, USA
\newline
% ...
$^{ 22}$Department of Physics and Astronomy, Tel Aviv University,
Tel Aviv 69978, Israel
\newline
$^{ 23}$International Centre for Elementary Particle Physics and
Department of Physics, University of Tokyo, Tokyo 113-0033, and
Kobe University, Kobe 657-8501, Japan
\newline
% ...
$^{ 27}$University of British Columbia, Department of Physics,
Vancouver BC V6T 1Z1, Canada
\newline
% ...
\bigskip

\cleardoublepage

\section{Introduction}
\label{sec:Intro}

Doubly charged Higgs bosons (\dbHpmpm) appear  
in theories beyond the Standard Model~\cite{ref:hpp_old}. Many authors
have pointed out that these doubly charged scalar particles occur naturally in
the left-right symmetric models which allow small 
neutrino masses~\cite{ref:hpp_leftright}, and stress the importance of 
looking for their existence~\cite{ref:hpp_importance}. 
Supersymmetric left-right models 
in which the $SU(2)_\mathrm{R}$ gauge
symmetry is broken by triplet Higgs fields do not conserve 
baryon and lepton numbers.
They also yield a natural embedding of 
a large mass for the right-handed Majorana neutrino needed for the
implementation of the see-saw mechanism for small neutrino masses. 
Recently, it has pointed out that such models can lead to 
``light''  doubly charged Higgs particles, in the 100~GeV/$c^2$
mass range~\cite{ref:hpp_theory}. 
An important 
feature of the doubly charged Higgs bosons
is that they couple, at tree level, only to charged leptons and to 
other Higgs and gauge bosons, in order to
conserve electric charge. Thus doubly charged Higgs bosons
could be copiously produced in pairs via
$\ee \rightarrow \mdbHpp \mdbHmm$
at LEP II with a differential cross section at tree level depending on 
the doubly charged Higgs boson mass and the centre-of-mass energy of the
colliding beams. In left-right symmetric models, two Higgs triplets
arise, conventionnaly labelled left and right handed. The cross section
of $\ee \rightarrow$ \dbHLpp \dbHLmm\ is not equal to that for 
$\ee \rightarrow$ \dbHRpp \dbHRmm. The events are characterized by the
$\sin^2 \theta$ dependence, where $\theta$ is the angle of the \dbHpp\ with 
respect to 
the electron beam axis, anticipated for the production
of a pair of spin 0 particles via the s-channel Z or $\gamma$
exchange.

The partial widths of the dominant decay modes,
    $\mdbHpp \rightarrow \ell^+ \ell^+$ and
    $\mdbHmm \rightarrow \ell^- \ell^-$, have been calculated by 
Huitu and collaborators~\cite{ref:hpp_formulae} 
and by Swartz~\cite{ref:hpp_limits}. 
For instance for H decaying into $\tau\tau$ it is, at tree level:
\begin{eqnarray}
\Gamma_{\tau \tau} 
    (\mdbHpmpm \rightarrow \tau^{\pm} \tau^{\pm}) = 
    \frac{h^{2}_{\tau \tau}}{8 \pi} \: M_{\mathrm{H}} \,
    \left( 1-\frac{2 m^{2}_{\tau}}{M^{2}_{\mathrm{H}}} \right)
    \left( 1-\frac{4 m^{2}_{\tau}}{M^{2}_{\mathrm{H}}} \right)^{1/2}
\label{decaywidth}
\end{eqnarray}
where 
   $M_{\mathrm{H}}$ is the mass of the doubly charged Higgs boson,
   $m_{\tau}$ is the mass of the tau lepton and
   $h_{\tau \tau}$ is the unknown H$\tau\tau$ Yukawa coupling constant.
If $M_{\mathrm{H}} \gg m_{\tau}$, values 
of $h_{\tau\tau} \gtrsim 10^{-7}$  
can ensure the fast decay of the Higgs particle ($\lesssim 10^{-11}$~s). 
Observation of a four lepton final state would provide a 
clean signature for the identification of the event. 
On the other hand for values of $h_{\tau\tau}$ smaller 
than about $10^{-7}$ long lived doubly charged Higgs bosons 
could be detected as events with large impact parameter tracks, 
kinked tracks, or stable doubly charged massive particles. OPAL has 
searched for charged massive particles at centre-of-mass energies up 
to 183~GeV~\cite{ref:opal_heavy}.

Previous searches for \dbHpmpm\ pair production have been performed
with the MARKII~\cite{ref:hpp_markII} and 
the OPAL~\cite{ref:hpp_opal} detectors
using data collected near the Z peak and have 
excluded the existence of the \dbHpmpm\ with mass less than 45.6~GeV/c$^{2}$ 
and for values of $h_{\tau \tau}$ that extend down to zero, apart 
from a small band around $h_{\tau \tau} \sim 10^{-7}$.

Significant couplings to electrons or muons 
are not likely as there are indirect 
constraints~\cite{ref:hpp_limits,ref:hpp_limits_2}
on the decays 
of $\mdbHpp \rightarrow \mathrm{e}^{+} \mathrm{e}^{+}$  
from high energy Bhabha scattering, 
of $\mdbHpp \rightarrow \mu^{+} \mu^{+}$ 
from the absence of muonium -- anti-muonium transitions, 
and of $\mdbHpp \rightarrow \mathrm{e}^{+} \mu^{+}$ 
from limits on the flavour changing decay 
$\mu^{-} \rightarrow \mathrm{e}^{+} \mathrm{e}^{-} \mathrm{e}^{-}$.
In this paper, however, we will present our results based on general 
searches for 
$\ee \rightarrow$ four leptons with missing energy (A), and four leptons 
without missing energy (B). We have considered 
the following channels:
(A) $\mdbHpp \mdbHmm \rightarrow 
 \tau^{+} \tau^{+} \tau^{-} \tau^{-}$,
$\mathrm{e}^{+} \mathrm{e}^{+} \tau^{-} \tau^{-}$ 
and
$\mu^{+} \mu^{+} \tau^{-} \tau^{-}$,
and (B) $\mdbHpp \mdbHmm \rightarrow
 \mathrm{e}^{+} \mathrm{e}^{+} \mathrm{e}^{-} \mathrm{e}^{-}$, 
$\mu^{+} \mu^{+} \mu^{-} \mu^{-}$
and 
$\mathrm{e}^{+} \mathrm{e}^{+} \mu^{-} \mu^{-}$.

The data set
is from \epair\ collisions recorded with the OPAL detector at LEP, 
and is composed of events at centre-of-mass
energies between $\roots=$ 189~GeV and 209~GeV with a total integrated
luminosity of about 614~pb$^{-1}$.
A complete description of the OPAL detector can be found 
in~\cite{ref:OPAL-detector,ref:SW,ref:TE}.

\section{Monte Carlo Simulation}
\label{sec:MC}

Monte Carlo samples are used to model the 
pair production of doubly charged Higgs bosons as well as to estimate 
the expected background due to \smp.

The simulation of the signal events with one non-zero 
$h_{\ell \ell '}$ coupling at a time and with zero lifetime
has been done with the Monte Carlo 
generator PYTHIA~\cite{ref:PYTHIA} modified according 
to~\cite{ref:hpp_formulae}. \dbHpp \dbHmm\ events have been generated
for various mass points ranging from 45~GeV/$c^{2}$ 
to half the centre-of-mass energy, $\roots$.
At each point on the ($M_{\mathrm{H}}$,$\roots$) plane, 
10000 events for each of the six
$\mdbHpmpm \rightarrow \ell^{\pm} \ell^{\pm}$ decays  
have been generated.

The main sources of background arise from 
\sm\ four-fermion final states ($\ee \rightarrow 4\mathrm{f}$) 
and a small contribution from two-photon 
($\ee \rightarrow \gamma\gamma \rightarrow$ hadrons, $\ell^+ \ell^-$) 
and two-fermion ($\ee \rightarrow \mathrm{Z} \rightarrow$ ff) processes. 
For two-photon processes, the PHOJET~\cite{ref:PHOJET},
PYTHIA~\cite{ref:PYTHIA}, and HERWIG~\cite{ref:herwig} generators have
been used to simulate hadronic final states.
The Vermaseren~\cite{ref:VERMASEREN} generator has been
used to estimate the background contribution from all 
two-photon $\ee \ell^+ \ell^-$ final states.
All other four-fermion final states have been simulated 
with grc4f~\cite{ref:grace4f}.
For the two-fermion final states, \mbox{BHWIDE}~\cite{ref:BHWIDE} was used for
the ee$(\gamma)$ final state, and KORALZ~\cite{ref:KORALZ} and 
KK2f~\cite{ref:KK2f} for the $\mu \mu$ and $\tau \tau$ states. 
The multi-hadronic events, 
${\rm qq}(\gamma)$, were simulated using PYTHIA and 
KK2f. 

All events were processed
through the full simulation of the OPAL detector~\cite{ref:GOPAL},
with the same analysis chain being applied to simulated events
and to data.

\section{Data Analysis}
\label{sec:Analysis}

The final states resulting from $\ee \rightarrow \mdbHpp \mdbHmm$ processes 
followed by $\mdbHpmpm \rightarrow \ell^{\pm} \ell^{\pm}$ decays
consist of four leptons. 
Two different analyses are applied depending on the decay assumed:
for $\mdbHpmpm \rightarrow 
      \tau^{\pm} \mathrm{e}^{\pm}$, 
     $\tau^{\pm} \mu^{\pm}$
 or  $\tau^{\pm} \tau^{\pm}$, a selection for four lepton candidates 
plus missing energy (resulting from the decays of the tau leptons) is applied, 
while for $\mdbHpmpm \rightarrow
      \mathrm{e}^{\pm} \mathrm{e}^{\pm}$,
     $\mu^{\pm} \mu^{\pm}$
 or  $\mathrm{e}^{\pm} \mu^{\pm}$, a selection for four leptons 
(electrons or muons) without missing energy is used. Although lepton 
identification is used in the event selection, it is not used 
for splitting the selected events into the different channels within
the two analyses. 

The analyses are similar to the
R-parity violation search~\cite{ref:opal_rpv_gauginos, 
ref:opal_rpv_sfermions} that considers four leptons with or without 
missing energy topologies. These event selection procedures 
are described 
in detail in \cite{ref:opal_rpv_gauginos, 
ref:opal_rpv_sfermions} and are only briefly outlined below. The cut values 
used in the present analysis are optimized for the searches for 
$\mdbHpmpm \rightarrow \tau^{\pm} \tau^{\pm}$ (in the missing energy 
analysis) and for
$\mdbHpmpm \rightarrow \mathrm{e}^{\pm} \mathrm{e}^{\pm}$ (in the 
no missing energy analysis). The selections are then applied as such 
for the other channels considered, and the efficiencies are calculated
separately for each channel. 

The visible energy and the momentum of the event are calculated using 
the method described in~\cite{ref:opal_calc_momentum}.
A preselection is applied, consisting of
\begin{itemize}
  \item Data quality requirements as described in~\cite{ref:opal_quality};
  \item Vetoes on the energy deposited in the forward detectors 
    to reduce background due to two-photon events
    and interactions of beam particles with the beam pipe or residual gas:
    the energy deposited in each silicon-tungsten forward calorimeter 
    and in each forward detector has to be less than 5~GeV
    (these detectors are located in the forward region, with polar 
    angle~\footnote{A right-handed coordinate system is adopted,
      where the $x$-axis points to the centre of the LEP ring,
      and positive $z$ is along  the electron beam direction.
      The angles $\theta$ and $\phi$ are the polar and azimuthal angles,
      respectively.} $|\cos \theta|>0.99$, surrounding the beam pipe);
    there should be no signal in the MIP plug scintillators~\footnote{The MIP plug 
      scintillators~\cite{ref:TE} are an array of 
      thin scintillating tiles with embedded wavelength shifting fibre
      readout which have been installed to improve the hermiticity of 
      the detector. They cover the polar angular range between 43 and
      200~mrad.}\@.
  \item A requirement that at least four and no more than 10 
    tracks are reconstructed.
\end{itemize}

After the preselection, a series of cuts is applied as follows. 

For the missing energy selection:
\begin{description}
  \item[(A1)]
    In order to reduce the background from two-photon processes 
    and from radiative return events 
    ($\ee \rightarrow \mathrm{Z}\gamma$) with the $\gamma$ escaping 
    into the beam pipe:
    \begin{itemize}
      \item The event's transverse momentum, \pt, 
        is required to be larger than $0.035 \roots /c$. 
      \item The polar angle, $\thmiss$, of the missing momentum 
        direction is required to satisfy $\cosmiss < 0.9$.
      \item The event's longitudinal momentum, \pz, is required to be smaller 
        than $0.25 \roots /c$.
      \end{itemize}
  \item[(A2)]
    The visible energy, $\Evis$, of the event is 
    required to be between $0.35 \roots$ and $0.85 \roots$.
  \item[(A3)]
    There have to be at least three tracks with a transverse momentum 
    with respect to the beam axis larger than 1.5~GeV/$c$.
  \item[(A4)]
    There have to be at least three well identified isolated 
    leptons~\cite{ref:opal_rpv_sfermions} 
    (e, $\mu$, $\tau$),
    each with a transverse momentum larger than 1.5~GeV/$c$.
  \item[(A5)]
    The total leptonic momentum, $\plept$, defined as the scalar sum of the 
    momenta of all identified leptons, is required to be greater 
    than 0.40 $\Evis /c$.
\end{description}

For the no missing energy selection:
\begin{description}
  \item[(B1)]
    The polar angle, $\thmiss$, of the missing momentum 
    direction has to satisfy $\cosmiss < 0.9$, in order to reject events
    in which a possible missing momentum points towards the beam pipe. 
  \item[(B2)]
    $\Evis$ is required to be larger than $0.90 \roots$.
  \item[(B3)]
    There have to be at least three isolated tracks
    each with a transverse momentum larger than 1.5~GeV/$c$.
  \item[(B4)]
    There have to be at least three well identified isolated 
    leptons~\cite{ref:opal_rpv_sfermions} 
    (e, $\mu$, $\tau$),
    each with a transverse momentum larger than 1.5~GeV/$c$.
  \item[(B5)]
    The total leptonic momentum is required to be greater 
    than 0.60 $\Evis /c$.
  \item[(B6)]
    In order to select events with a pair of particles with identical masses, 
    the following procedure is applied.
    Pairs are formed using the four most energetic tracks, 
    and the invariant
    mass is computed for each pair. Events are selected if 
    one of the three possible pairings satisfies 
    $ |m_{i,j} - m_{k,l}|/(m_{i,j} + m_{k,l}) < 0.3 $, where 
    $m_{i,j}$ is the invariant mass of the pair $(i,j)$. Only pairs 
    with invariant masses $m_{i,j}$ greater than 20~GeV/$c^{2}$ are 
    used in the computation.
\end{description}

The numbers of events remaining after each cut are listed in 
Tables~\ref{tab:cut_flow_miss} and~\ref{tab:cut_flow_nomiss}.
The poor agreement 
between data and Monte Carlo expectation in the early stages of the 
selections is
due to the beam-related backgrounds 
and incomplete modelling of two-photon processes.
When the background from these processes has been 
effectively reduced, the agreement between data and Monte Carlo is 
satisfactory.

\begin{figure}[]
  \centering
  \epsfig{file=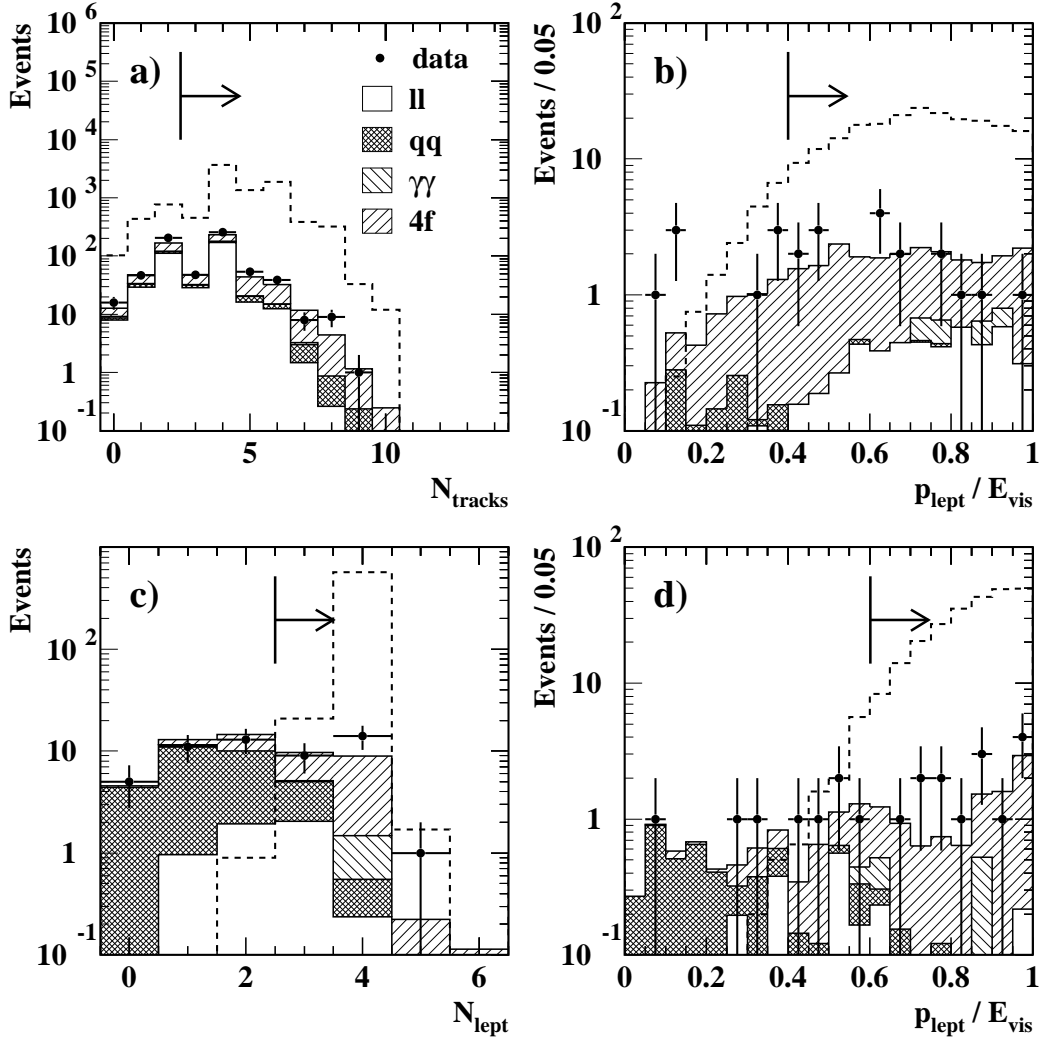,width=15.5cm} 
  \caption[]{\sl
    Distributions of some essential observables used in the selections: 
    a) The number of tracks with transverse momentum larger than 1.5~GeV/$c$, 
    $\ntracks$, after cut (A2) in the 
    missing energy selection.
    b) The total leptonic momentum divided by the total visible 
    energy of the event, $\plept/\Evis$, after cut (A4) 
    in the missing energy selection.
    c) The number of well identified isolated leptons, $\nlept$,
    after cut (B3) in the no missing energy selection.
    d) $\plept/\Evis$ after cut (B4) in the no missing energy selection.
    The data are shown with error bars and distributions from the 
    background processes are shown as solid line histograms: 
    dilepton events (open area), 
    multihadronic events (double hatched area), 
    two-photon processes (negative slope hatching area), 
    and four-fermion processes (positive slope hatching area).
    The arrows, pointing into the accepted regions, show where 
    the analysis cuts are applied. The dashed line histograms show 
    the predictions for a doubly-charged Higgs boson signal 
    with $M_{\mathrm{H}} = 90$~GeV/$c^{2}$ at $\roots = 206$~GeV 
    with 100\% decay into $\tau \tau$ for a) and b) 
    and 100\% decay into ee for c) and d). The normalisations of the signal 
    histograms are arbitrary.
    }
  \label{fig:distributions}
\end{figure}

Figure~\ref{fig:distributions} shows the distributions of
some of the variables used in the selections 
for the data and the background simulation (the contributions 
from all centre-of-mass energies are added).
For the missing energy selection, 
figure~\ref{fig:distributions}a) shows the distributions of 
the number of tracks with a transverse momentum 
with respect to the beam axis larger than 1.5~GeV/$c$
after cut (A2) has been applied.
The distributions for a signal sample for 
$M_{\mathrm{H}} = 90$~GeV/$c^{2}$ and 
$\mdbHpmpm \rightarrow \tau^{\pm} \tau^{\pm}$ 
at $\roots = 206$~GeV are also shown with arbitrary normalisation.
Similarly, figure~\ref{fig:distributions}b) shows the distributions 
of the total leptonic momentum divided by the visible energy of the event
after all but cut (A5) have been applied.

For the no missing energy selection, 
figure~\ref{fig:distributions}c) shows the distributions of 
the number of well identified isolated leptons after cut (B3) 
and figure~\ref{fig:distributions}d) the distributions
of the total leptonic momentum divided by the visible energy of the event
after cut (B4). 
The distributions for a signal sample for 
$M_{\mathrm{H}} = 90$~GeV/$c^{2}$ and 
$\mdbHpmpm \rightarrow \mathrm{e}^{\pm} \mathrm{e}^{\pm}$ 
at $\roots = 206$~GeV are also shown with arbitrary normalisation.

The signal detection efficiencies are given in Table~\ref{tab:results}.
For values of $h_{\ell \ell}$ smaller than approximately $10^{-7}$, 
the efficiency to detect $\mdbHpmpm \rightarrow \ell^{\pm} \ell^{\pm}$ 
decreases due to the long decay length of the $\mdbHpmpm$. At very 
small $h_{\ell \ell}$ ($10^{-8}$ to $10^{-9}$) the $\mdbHpmpm$ could 
traverse the tracking chambers of the detector without decaying.
We assume zero detection efficiency for values of $h_{\ell \ell}$ 
smaller than $10^{-7}$.

\section{Inefficiencies and Systematic Uncertainties}
\label{sec:Syst}

The following systematic errors on the signal detection efficiencies
have been considered:
the statistical uncertainty on the determination of the efficiency from the 
Monte Carlo simulation (less than 1\% absolute uncertainty);
the systematic uncertainty on the measurement of the integrated 
luminosity (0.20\% to 0.24\%); 
and the errors due to the modelling of the physics
variables (between 0.5\% and 4.5\% absolute uncertainty). 
These last errors are determined by shifting each cut by 
the maximum amount such that distributions from high 
statistics Monte Carlo samples remain in agreement with 
the data.

For the expected number of background events, 
an uncertainty due to Monte Carlo statistics (4\% to 8\%) and  
a systematic error due to the modelling of the physics
variables (5\% to 7\%) have been taken into account.

The total systematic uncertainty was calculated by summing in quadrature 
the individual errors. Correlations between systematic uncertainties at 
different centre-of-mass energies have been taken into account.

In addition to effects included in the detector simulation, a relative
efficiency loss of 1.5\% to 2.9\% arises from beam-related
background in the silicon-tungsten forward calorimeter and 
in the forward detector. This
is estimated using random beam crossing events.

\section{Interpretation}
\label{sec:Interp}

As shown in Table~\ref{tab:results} 
the observed numbers of events selected by the 
missing energy and the no missing energy
selections (17 and 10) are consistent with the numbers expected  
from Standard Model processes (25.19 $\pm$ 0.92 $\pm$ 1.06 and 6.57 $\pm$
0.50 $\pm$ 0.31). 
No evidence for \dbHpp \dbHmm\ production has therefore been observed 
and limits
on the doubly charged Higgs boson production cross-section and mass have
been derived.

\begin{figure}[]
  \centering
  \epsfig{file=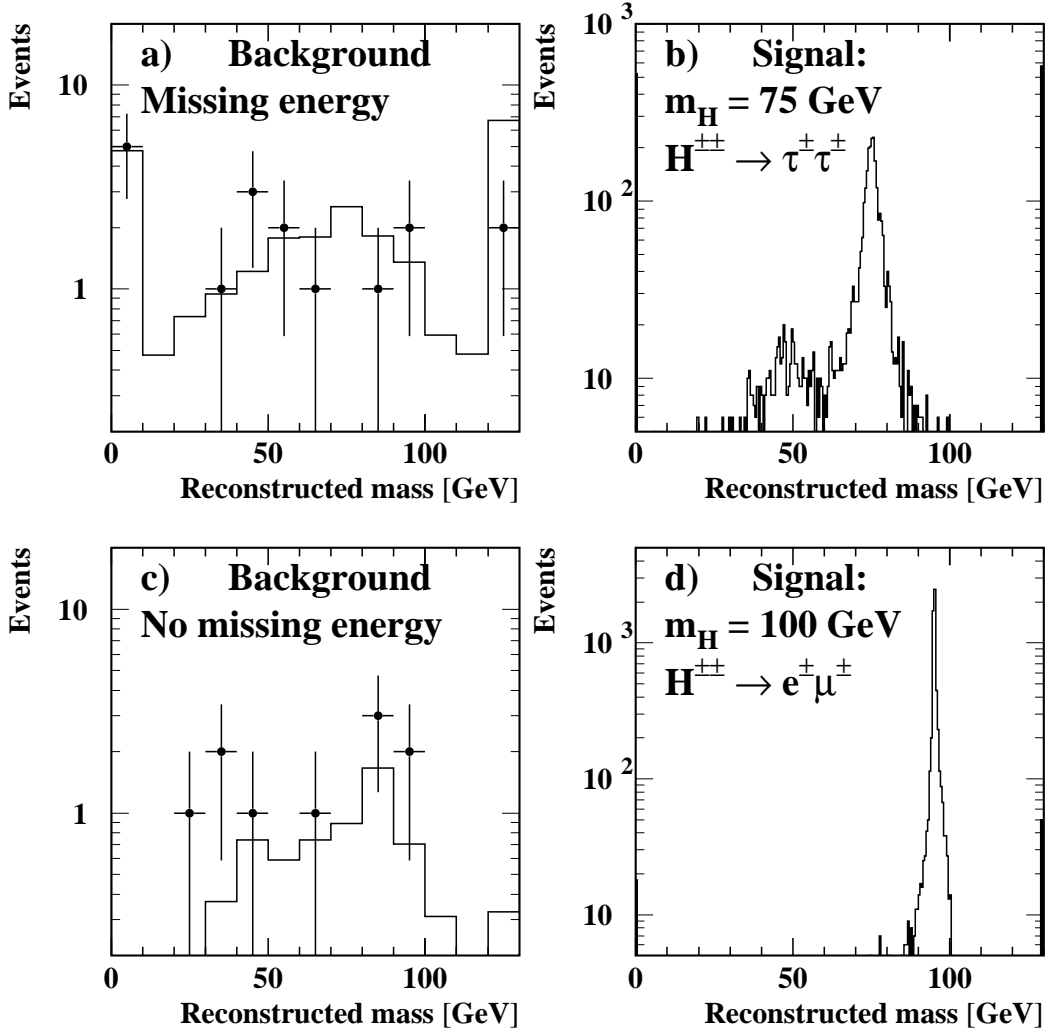,width=15.5cm} 
  \caption[]{\sl
    Distributions of the reconstructed Higgs boson mass for 
    a) background (solid-line histogram) and data (points with error bars) and
    b) a signal with $M_{\mathrm{H}} = 75$~GeV/$c^{2}$ with 100\% 
    decay into $\tau \tau$ 
    at $\roots = 206$~GeV, 
    in the missing energy selection, and 
    c) background (solid-line histogram) and data (points with error bars) and
    d) a signal with $M_{\mathrm{H}} = 100$~GeV/$c^{2}$ with 100\% decay 
    into ee, 
    at $\roots = 206$~GeV, in the no missing energy selection.
    } 
  \label{fig:recmass}
\end{figure}

For each selected event, a reconstructed mass is computed. 
Figure~\ref{fig:recmass} shows data and background as well as typical 
signal reconstructed mass distributions
for the e$^+\mu^+$e$^-\mu^-$ and $\tau^+\tau^+\tau^-\tau^-$ states.
The momenta of
the four lepton candidates are constrained so that the total energy
is the centre-of-mass energy, and the total momentum adds to zero.
For events with unambiguous charge assignments (two positive lepton
candidates and
two negative lepton candidates), the charge is used to assign leptons to
the
hypothesized doubly-charged Higgs bosons.  The reconstructed mass is then
the average of the invariant masses reconstructed for the two boson
hypotheses in the event.  For events for which the charge assignment is
ambiguous (three or more lepton candidates of the same sign), the
assignment
of lepton candidates to hypothesized Higgs bosons is chosen to minimize
the
difference between the two reconstructed masses of the Higgs bosons, and
the average reconstructed mass of the two bosons is used as the event
reconstructed mass.  In the channel without missing energy, the
reconstructed mass resolution is characterized by two components: a core
of width 400~MeV/$c^2$ with approximately 96\% of the signal, and a tail 
of width 3~GeV/$c^2$, with approximately 4\% of the signal. In the channel
with missing energy, for the $\tau^+\tau^+\tau^-\tau^-$ final state, for
example,
approximately 40\% of the signal is concentrated within a peak of width
2.5~GeV/$c^2$, while the remainder has a long tail.  
The first bin of the histograms collects events which fail to have 
a reconstructed mass (e.g. missing a lepton candidate), and the last 
bin contains all overflows (events with a reconstructed mass larger 
than 130~GeV/$c^2$.

The distributions of the reconstructed masses of signal Monte Carlo
events 
for various test masses are formed, with bin widths of 500~MeV/$c^2$.
To obtain the expected signal distribution for an arbitrary test mass,
the available signal histograms are interpolated~\cite{ref:readinterp}.
Histograms of the reconstructed masses of Standard Model Monte Carlo events,
averaging over 22~GeV/$c^2$ wide mass intervals, are used as the background.

A likelihood ratio method \cite{ref:likelihood} has been used to 
determine an upper limit for the production cross-section. 
This method combines the results obtained at different centre-of-mass
energies, taking into account the number of candidates and their 
reconstructed masses, the signal
detection efficiencies, the expected number of background 
events and the distributions of the reconstructed masses for 
background and for signal events.
The confidence levels are computed by binning the data in the
reconstructed mass separately for the different centre-of-mass 
energies\footnote{When calculating limits, cross-sections at 
different $\roots$ are estimated by weighting by $\beta^3/s$, 
where $\beta$ is 
$p_{\mdbHpmpm} / \mathrm{E}_{\mathrm{beam}} $ },
and combining them.
Systematic uncertainties on the efficiencies and on the number of 
expected background events have been 
taken into account in the limit calculation 
according to~\cite{ref:cousins}.

\begin{figure}[]
  \centering
  \epsfig{file=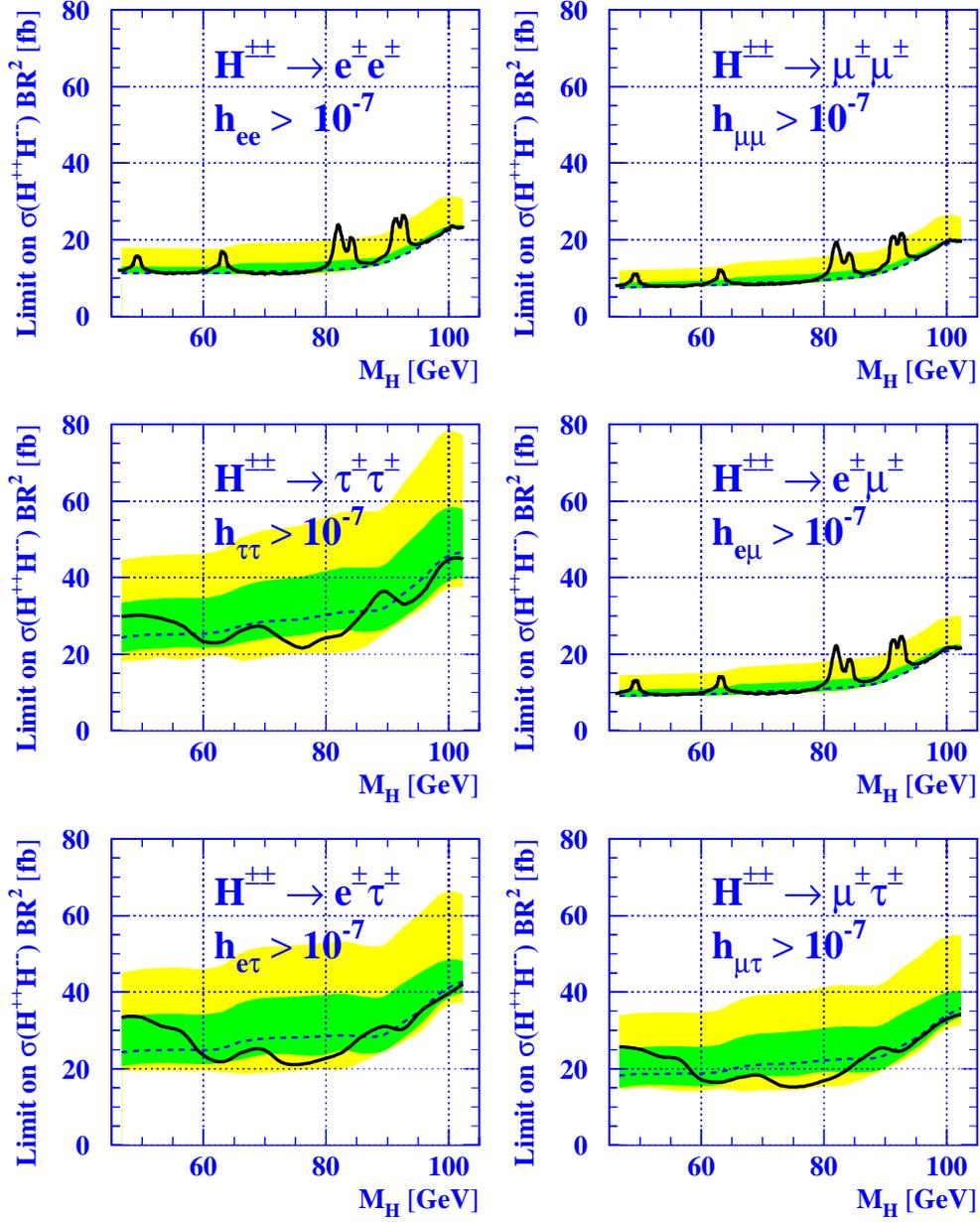,width=14.5cm} 
  \caption[]{\sl
    The 95\% confidence level 
    upper limits on $\sigma(\ee \rightarrow \dbHpp \dbHmm) 
    \times BR^{2}(\dbHpmpm \rightarrow \ell^{\pm} \ell^{'\pm})$ for each of
    the six possible leptonic decays at $\roots = 206$~GeV.
    On each plot, the solid line shows the observed limit, 
    while the dotted line 
    shows the limit expected for background only. The dark/light 
    shaded bands around the background 
    expectation represent the $\pm1/\pm2$ standard deviation spread.
    } 
  \label{fig:cross_limit_all}
\end{figure}

Within supersymmetric left-right symmetric models, this results in
limits on $\sigma(\ee \rightarrow \dbHpp \dbHmm) 
          \times BR^{2}(\dbHpmpm \rightarrow \ell^{\pm} \ell^{'\pm})$ 
given as a function of the \dbHpmpm\ mass, where 
$BR(\dbHpmpm \rightarrow \ell^{\pm} \ell^{'\pm})$ 
is the branching ratio of the decay 
$\dbHpmpm \rightarrow \ell^{\pm} \ell^{'\pm}$. 
Figure~\ref{fig:cross_limit_all} shows the 95\% confidence level upper
limit on $\sigma(\ee \rightarrow \dbHpp \dbHmm) 
          \times BR^{2}(\dbHpmpm \rightarrow \ell^{\pm} \ell^{'\pm})$ 
for each of the six possible leptonic decays assuming conservatively that
the efficiency for all other decays is negligible.

\begin{figure}[]
  \centering
  \epsfig{file=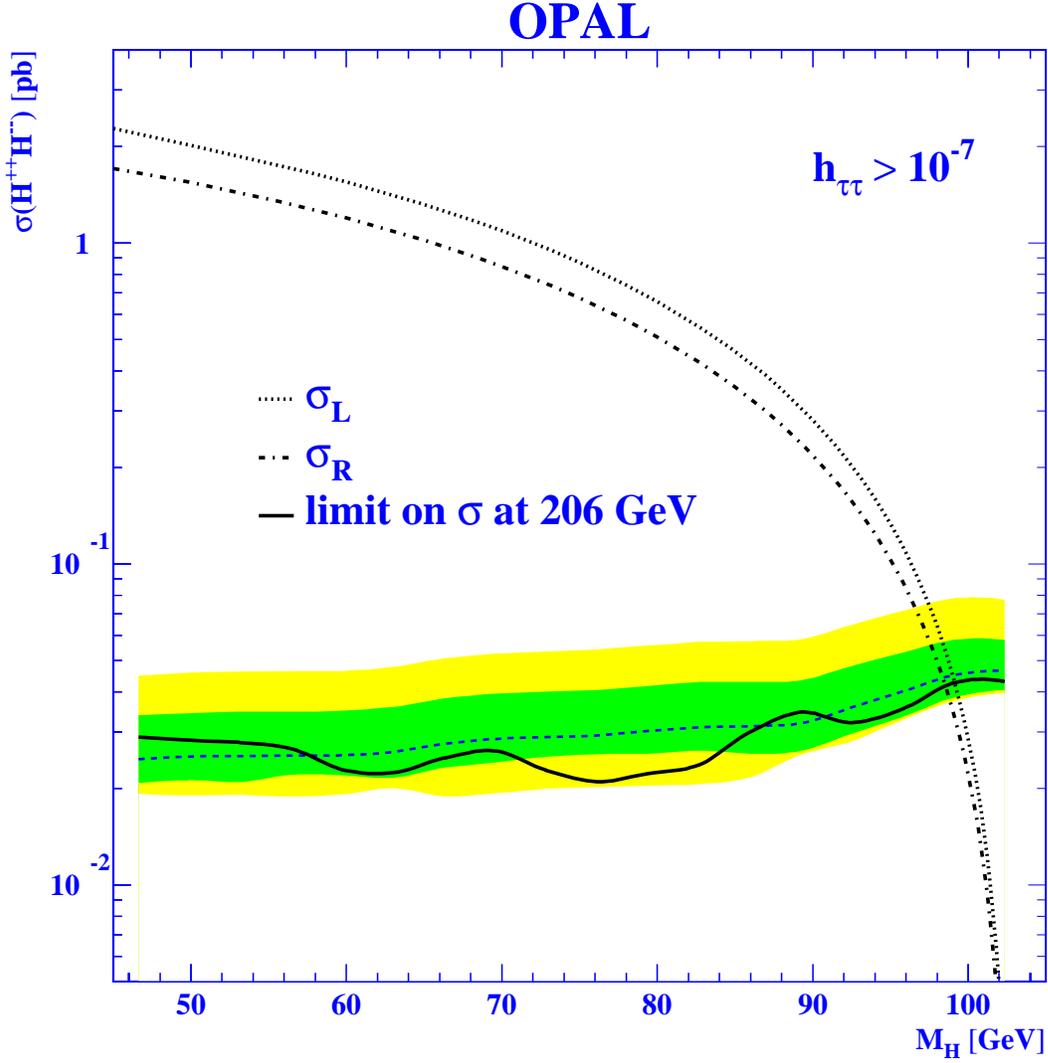,width=15.5cm} 
  \caption[]{\sl
    The solid line shows the 95\% confidence level 
    upper limit on the \dbHpmpm\ pair 
    production cross-section at $\roots = 206$~GeV
    assuming 100\% branching ratio for the decay of \dbHpmpm\
    into $\tau^{\pm} \tau^{\pm}$. 
    The dashed line shows the limit expected for background only. 
    The dark/light shaded bands around the background 
    expectation represent the $\pm1/\pm2$ standard deviation spread.
    The dotted and dash-dotted lines show the expected production 
    cross-sections of \dbHLpp \dbHLmm\ and \dbHRpp \dbHRmm\ 
    in left-right symmetric models. All data collected at $\roots$ between 
    189~GeV and 209~GeV have been combined and the production cross-section
    $\sigma$ has been assumed to vary linearly with $\beta^3/s$.
    } 
  \label{fig:cross_limit_comb}
\end{figure}

As an example, 
Figure~\ref{fig:cross_limit_comb} shows the 95\% confidence level upper
limit on the cross-section for pair production of \dbHpmpm\ followed by a
decay into $\tau^{\pm} \tau^{\pm}$ at $\roots = 206$ GeV, obtained 
from the data collected at centre-of-mass energies 
between 189~GeV and 209~GeV. Comparing this limit with the expected 
cross-sections
for left-handed \dbHLpmpm\ and right-handed \dbHRpmpm\ pair 
production~\cite{ref:hpp_formulae}, $\sigma_{\mathrm{L}}$ and 
$\sigma_{\mathrm{R}}$,
yields 95\% C.L. lower mass limits of 
99.0~GeV/$c^{2}$ and 98.5~GeV/$c^{2}$, respectively, assuming 
a 100\% decay into $\tau^{\pm} \tau^{\pm}$.
As listed in Table~\ref{tab:mass_limits} the left-handed \dbHLpmpm\ 
mass limits for all channels taken independently lie between 
99.0~GeV/$c^{2}$ (for 100\% decay into $\tau^{\pm} \tau^{\pm}$) 
and 100.5~GeV/$c^{2}$ (for $\mu^{\pm} \mu^{\pm}$).
For right-handed \dbHRpmpm\, the mass limits lie between 
98.5~GeV/$c^{2}$ and 100.1~GeV/$c^{2}$.

\section{Conclusion}
\label{sec:Concl}

The OPAL data sets at centre-of-mass energies between 189~GeV and 209~GeV,
corresponding to a total luminosity of about 614~pb$^{-1}$, have been
searched for evidence of the reaction $\ee \rightarrow \mdbHpp \mdbHmm$
followed by the decays $\mdbHpmpm \rightarrow \ell^{\pm} \ell^{'\pm}$ in
R-parity conserving supersymmetric left-right symmetric models. 
No significant excess of events has been observed in the data.
Production cross-section limits have been derived 
for values of the 
Yukawa coupling constant $h_{\ell \ell^{'}}$ larger than $10^{-7}$.
A lower mass limit of 98.5~GeV/$c^{2}$ at the 95\% confidence level 
has been obtained for Higgs bosons decaying via a single channel 
$\mdbHpmpm \rightarrow \ell^{\pm} \ell^{'\pm}$ with 100\% branching ratio.

\section{Acknowledgements}

We particularly wish to thank the SL Division for the efficient operation
of the LEP accelerator at all energies
 and for their close cooperation with
our experimental group.  We thank our colleagues from CEA, DAPNIA/SPP,
CE-Saclay for their efforts over the years on the time-of-flight and trigger
systems which we continue to use.  
We also 
gratefully thank R.N.~Mohapatra for valuable support on
theoretical issues.

In addition to the support staff at our own
institutions we are pleased to acknowledge the  \\
Department of Energy, USA, \\
National Science Foundation, USA, \\
Particle Physics and Astronomy Research Council, UK, \\
Natural Sciences and Engineering Research Council, Canada, \\
Israel Science Foundation, administered by the Israel
Academy of Science and Humanities, \\
Minerva Gesellschaft, \\
Benoziyo Center for High Energy Physics,\\
Japanese Ministry of Education, Science and Culture (the
Monbusho) and a grant under the Monbusho International
Science Research Program,\\
Japanese Society for the Promotion of Science (JSPS),\\
German Israeli Bi-national Science Foundation (GIF), \\
Bundesministerium f\"ur Bildung und Forschung, Germany, \\
National Research Council of Canada, \\
Research Corporation, USA,\\
Hungarian Foundation for Scientific Research, OTKA T-029328, 
T023793 and OTKA F-023259,\\
Fund for Scientific Research, Flanders, F.W.O.-Vlaanderen, Belgium.\\

\cleardoublepage



\newpage
\begin{table}[]
  \centering
  \begin{tabular}{|l||r||r||r|r|r|r||r|}
    \hline
    Cut  & data & total bkg & $\ellell (\gamma)$& $\qq (\gamma)$ & 
    `$\gamma \gamma$' & 4-f & signal \\
    \hline
    \hline
Pres.&   80543 &    51158.7 &     1667.5 &     1210.5 &    47580.5 &      700.2 &       86.8 \% \\
A1   &    1550 &     1318.1 &      676.6 &      194.6 &      156.5 &      290.4 &       73.9 \% \\
A2   &     684 &      601.0 &      379.3 &       28.3 &        6.8 &      186.5 &       62.7 \% \\
A3   &     417 &      375.0 &      230.3 &       17.7 &        3.0 &      123.9 &       54.0 \% \\
A4   &      25 &       30.0 &        5.6 &        0.5 &        0.9 &       22.9 &       47.7 \% \\
A5   &      17 &       25.2 &        5.3 &        0.1 &        0.9 &       18.9 &       44.5 \% \\
     &         & $\pm$  1.4 & $\pm$  0.4 & $\pm$  0.1 & $\pm$  0.5 & $\pm$  1.2 & $\pm$  1.0 \% \\
    \hline
  \end{tabular}
  \caption[]{
    \protect{\parbox[t]{12cm}{
        The remaining numbers of events
        after each cut of the missing energy selection 
        for various background processes normalised to 614~pb$^{-1}$
        are compared with data collected at $\roots$ between 189~GeV 
        and 209~GeV.
        Efficiencies for a simulated event sample of $\mdbHpp \mdbHmm$ with
        $\mdbHpmpm \rightarrow \tau^{\pm} \tau^{\pm}$ and 
        $M_{\mathrm{H}} = 90$~GeV/$c^{2}$ at $\roots = 206$~GeV are also given.
        }} 
    }
  \label{tab:cut_flow_miss}
\end{table}

\begin{table}[]
  \centering
  \begin{tabular}{|l||r||r||r|r|r|r||r|}
    \hline
    Cut  & data & total bkg & $\ellell (\gamma)$& $\qq (\gamma)$ & 
    `$\gamma \gamma$' & 4-f & signal \\
    \hline
    \hline
Pres.&   80543 &    51158.7 &     1667.5 &     1210.5 &    47580.5 &      700.2 &       84.7 \% \\
B1   &   20655 &    13878.8 &      909.9 &      271.8 &    12341.4 &      355.7 &       71.9 \% \\
B2   &     571 &      557.3 &      277.5 &      204.4 &        5.0 &       70.3 &       71.3 \% \\
B3   &      53 &       50.9 &        5.2 &       26.2 &        1.1 &       18.4 &       59.4 \% \\
B4   &      24 &       18.8 &        2.3 &        3.3 &        1.0 &       12.2 &       59.3 \% \\
B5   &      16 &       11.3 &        0.8 &        0.2 &        0.9 &        9.4 &       57.2 \% \\
B6   &      10 &        6.6 &        0.2 &        0.1 &        0.4 &        5.8 &       54.2 \% \\
     &         & $\pm$  0.6 & $\pm$  0.1 & $\pm$  0.1 & $\pm$  0.3 & $\pm$  0.5 & $\pm$  0.8 \% \\
    \hline
  \end{tabular}
  \caption[]{
    \protect{\parbox[t]{12cm}{
        The remaining numbers of events
        after each cut of the no missing energy selection 
        for various background processes normalised to 614~pb$^{-1}$
        are compared with data collected at $\roots$ between 189~GeV 
        and 209~GeV.
        Efficiencies for a simulated event sample of $\mdbHpp \mdbHmm$ with
        $\mdbHpmpm \rightarrow \mathrm{e}^{\pm} \mathrm{e}^{\pm}$ and 
        $M_{\mathrm{H}} = 90$~GeV/$c^{2}$ at $\roots = 206$~GeV are also given.
        }} 
    }
  \label{tab:cut_flow_nomiss}
\end{table}

\begin{table}[]
  \centering
  \begin{tabular}{|l||c|}
    \hline
    Decay                     & efficiencies (\%)        \\
    \hline
    \hline
    $\mathrm{e} \mathrm{e}$   & 42--61    \\
    $\mu \mu$                 & 66--71    \\
    $\tau \tau$               & 42--46    \\
    $\mathrm{e} \mu$          & 55--65    \\
    $\mathrm{e} \tau$         & 33--40    \\
    $\mu \tau$                & 44--47    \\
    \hline
  \end{tabular}
  \caption[]{
    \protect{\parbox[t]{12cm}{
        Ranges of signal detection efficiencies (in \%) for the different 
        \dbHpmpm\ masses (between 45~GeV/$c^{2}$ and 102~GeV/$c^{2}$)
        for the various decays.
        }} 
    }
  \label{tab:results}
\end{table}

\begin{table}[]
  \centering
  \begin{tabular}{|l||c|c|}
    \hline
 &  & \\
    Decay                     & \dbHLpmpm & \dbHRpmpm \\
 &  & \\
    \hline
    \hline
    $\mathrm{e} \mathrm{e}$   & 100.2 GeV/$c^{2}$ &  99.9 GeV/$c^{2}$ \\
    $\mu \mu$                 & 100.5 GeV/$c^{2}$ & 100.1 GeV/$c^{2}$ \\
    $\tau \tau$               &  99.0 GeV/$c^{2}$ &  98.5 GeV/$c^{2}$ \\
    $\mathrm{e} \mu$          & 100.4 GeV/$c^{2}$ & 100.0 GeV/$c^{2}$ \\
    $\mathrm{e} \tau$         &  99.3 GeV/$c^{2}$ &  98.7 GeV/$c^{2}$ \\
    $\mu \tau$                &  99.6 GeV/$c^{2}$ &  99.3 GeV/$c^{2}$ \\
    \hline
  \end{tabular}
  \caption[]{
    \protect{\parbox[t]{12cm}{
        Mass limits at the 95\% confidence level 
        for left-handed and right-handed doubly charged Higgs bosons \dbHpmpm.
        }} 
    }
  \label{tab:mass_limits}
\end{table}


\newpage

\end{document}